# A Machine Learning, Natural Language Processing Analysis of Youth Perspectives: *Key Trends and Focus Areas for Sustainable Youth Development Policies*

Kushaagra Gupta, International Development Researcher, Foothill College IHS (corresponding author)
kushgks@gmail.com
(510) 458 – 0658
12345 El Monte Road,
Los Altos Hills, CA, 94022

## Abstract

Investing in children and youth is a critical step towards inclusive, equitable, and sustainable development for current and future generations. Several international agendas for accomplishing common global goals emphasize the need for active youth participation and engagement for sustainable development. The 2030 Agenda for Sustainable Development emphasizes the need for youth engagement and the inclusion of youth perspectives as an important step toward addressing each of the 17 Sustainable Development Goals.[1] The Addis Ababa Action Agenda also recognizes the significance of investing in children and youth for achieving equitable sustainable development.[2] The aim of this study is to analyze youth perspectives, values, and sentiments towards issues addressed by each of the 17 Sustainable Development Goals through social network analysis using machine learning. Social network data collected during 7 major sustainability conferences aimed at engaging children and youth is analyzed using natural language processing techniques for sentiment analysis. This data is then categorized using a natural language processing text classifier trained on a sample dataset of social network data during the 7 youth sustainability conferences for a deeper understanding of youth perspectives in relation to each of the 17 Sustainable Development Goals. Machine learning identified demographic and location attributes and features are utilized in order to identify bias and demographic differences between ages, gender, and race among youth. Using natural language processing, the qualitative data collected from over 7 different countries in 3 languages are systematically translated, categorized, and analyzed, revealing key trends and focus areas for sustainable youth development policies. The obtained results reveal the general youth's depth of knowledge on sustainable development and their attitudes towards each of the 17 Sustainable Development Goals. The findings of this study serve as a guide toward better understanding the interests, roles, and perspectives of children and youth in achieving the goals of Agenda 2030.

---

[1] Desa, U. N. "Transforming our world: The 2030 agenda for sustainable development." (2016).

[2] United Nations. Department of Economic and Social Affairs. World youth report: Youth and the 2030 agenda for sustainable development. New York: United Nations Publications, 2018.

# I. INTRODUCTION

Following the Millennium Development Goals (MDGs), the international community introduced the Sustainable Development Goals (SDGs).[3] With 17 goals and 169 targets, the Sustainable Development Goals aim to build upon the Millennium Development Goals through various objectives. The main objectives include: ending poverty and hunger, protecting the planet from degradation, ensuring peace and prosperity, and implementing the 2030 Agenda for Sustainable Development through Global Partnership.[4] In order to achieve these objectives for sustainable development, investment in the youth is critical step.[5] Both the 2030 Agenda for Sustainable Development and the Addis Ababa Action Agenda of the Third International Conference on Financing for Development [6] identify investing in the youth as a critical step towards the realization of these goals and targets for sustainable development. Due to different nations' economic and cultural features, multiple definitions of youth in terms of upper age limit are accepted in practice. For this study, people between 13 and 25 years of age are categorized as youth. The significance of youth involvement and participation in reaching the Sustainable Development Goals has been highlighted among numerous relevant documents (i.e. Agenda 21, Addis Ababa Action Agenda, etc.), yet their current role, future positions, and general sentiment regarding each of the Sustainable Development Goals remain largely unrecognized.[7] The attitudes and perceptions of young people surrounding the Sustainable Development Goals are ambiguous, and youth engagement in policymaking is limited and relatively scarce. [8] However, it is evident that youth engagement is important for the success of the Sustainable Development Goals. [9]

---

[3] Sachs, Jeffrey D. "From millennium development goals to sustainable development goals." *The lancet* 379, no. 9832 (2012): 2206-2211.

[4] Desa, U. N. "Transforming our world: The 2030 agenda for sustainable development." (2016).

[5] Raikes, Abbie, Hirokazu Yoshikawa, Pia Rebello Britto, and Iheoma Iruka. "Children, Youth and Developmental Science in the 2015-2030 Global Sustainable Development Goals. Social Policy Report. Volume 30, Number 3." *Society for Research in Child Development* (2017).

[6] Agenda, Addis Ababa Action. "Addis Ababa Action Agenda of the third international conference on financing for development." *UN. development* 2 (2015).

[7] Petković, Jasna, Nataša Petrović, Ivana Dragović, Kristina Stanojević, Jelena Andreja Radaković, Tatjana Borojević, and Mirjana Kljajić Borštnar. "Youth and forecasting of sustainable development pillars: An adaptive neuro-fuzzy inference system approach." *PloS one* 14, no. 6 (2019).

[8] Martin, Shanetta, Karen Pittman, Thaddeus Ferber, and Ada McMahon. "Building Effective Youth Councils: A Practical Guide To Engaging Youth In Policy Making." In *Forum for youth investment*. Forum for Youth Investment. The Cady-Lee House, 7064 Eastern Avenue NW, Washington, DC 20012-2031, 2007.

[9] Petrova, Yu, D. Dzhioeva, and L. Edilsultanova. "The role of youth leadership in achieving sustainable development, environmental safety." In *AIP Conference Proceedings*, vol. 2442, no. 1, p. 060001. AIP Publishing LLC, 2021

## A. Youth Communication Channels

With over 1.2 billion people between the ages of 13 and 25, children and youth younger than 25 years of age constitute over 42% of the global population. [10] Young people have grown up in the digital era and are quick adopters to new innovation in technology. Access to the internet and mobile phones is spreading, particularly in regions with rapidly growing youth populations, such as the Sub-Saharan Africa. For a majority of the global youth population, the social media and the internet have become more popular news sources than newspapers. Digital media and social networks have impacted political participation among the youth across the globe. [11] [12] However, the internet's proliferation and growth of digital media have fractured the global media landscape. [13] As a result, communicating with broad audiences has grown increasingly challenging. Yet, the segmentation of media outlets allows communication with precise target audiences.

Social media has been an effective platform for garnering youth interest towards frameworks such as the 2030 Agenda for Sustainable Development and greater youth participation in policy-making. This study aims to utilize social media data to analyze youth trends based on age, race, and gender.

## B. Social Network Analysis

Mining, processing, and analyzing data from social networks have aided studies in politics, policymaking, transportation, economics, marketing, education, and disaster relief. [14] Public data mined from Twitter and Facebook have been utilized to monitor public health, [15] forecast stock prices [16] and reveal public opinions surrounding politics. [17] These studies demonstrate the value of mining data from public social networks on a national or global scale to augment or outperform conventional survey approaches in circumstances when they are impractical due to temporal or geographic size. Relatively recent advancements in machine learning and deep learning based natural language processing, such as word-level-embeddings, the Seq2Seq Framework, Continuous Bag-of-Words (CBOW), and recurrent neural networks, have transformed the way researchers analyze data. [18] Through transfer learning, researchers are able to utilize preexisting state of the art transformer models such as the Bidirectional Encoder Representations from Transformers and the Robustly Optimized Bidirectional Encoder Representations from Transformers Pre-training Approach without needing to develop a new natural language processing model from the beginning. Both the Bidirectional Encoder Representations from Transformers (BERT) and the Robustly Optimized Bidirectional Encoder Representations from Transformers Pre-training Approach (RoBERTa) allow researchers to utilize massive unlabeled datasets such as WikiText which consists of over 3.3 billion words.[19] Through BERT and other models based off of BERT-architecture, researchers have access to state-of-the-art models which can be fine-tuned for classification tasks such as sentiment analysis and question-and-answer tasks such as creating chatbots. BERT based models have been used to predict text in emails, quickly summarize legal documents, and translate over 100 languages.

---

[10] Ritchie, Hannah, and Max Roser. "Age structure." *Our World in Data* (2019).

[11] Kahne, Joseph, and Ellen Middaugh. "Digital media shapes youth participation in politics." *Phi Delta Kappan* 94, no. 3 (2012): 52-56.

[12] Kamau, Samuel C. "Democratic engagement in the digital age: youth, social media and participatory politics in Kenya." *Communicatio* 43, no. 2 (2017): 128-146.

[13] Albarran, Alan B. *The social media industries*. Edited by Alan B. Albarran. New York: Routledge, 2013.

[14] Gundecha, Pritam, and Huan Liu. "Mining social media: a brief introduction." *New directions in informatics, optimization, logistics, and production* (2012): 1-17.

[15] Paul, Michael J., Abeed Sarker, John S. Brownstein, Azadeh Nikfarjam, Matthew Scotch, Karen L. Smith, and Graciela Gonzalez. "Social media mining for public health monitoring and surveillance." In *Biocomputing 2016: Proceedings of the Pacific symposium*, pp. 468-479. 2016.

[16] Wang, Yaojun, and Yaoqing Wang. "Using social media mining technology to assist in price prediction of stock market." In *2016 IEEE International conference on big data analysis (ICBDA)*, pp. 1-4. IEEE, 2016.

[17] McGregor, Shannon C. ""Taking the temperature of the room" how political campaigns use social media to understand and represent public opinion." *Public Opinion Quarterly* 84, no. S1 (2020): 236-256.

Using natural language processing techniques on social media data, identifying keywords and overarching topics, classifying social media posts, and analyzing sentiment on a large population is possible. Additionally, through webpage traffic information, network analysis, and ethnicity-classification character-based recurrent neural networks, identifying demographics of social media users is possible. Demographic data extracted from social media platforms such as Twitter and Facebook have been used by researchers to research political polarization, racial segregation, public health, and urbanization. [20] Given the widely documented racial and socio-economic differences in global environmental movements, identifying the demographics of those engaged in sustainable development topics on social media in real-time may provide valuable insights into how diverse groups engage with sustainable development concerns.

## II. METHODOLOGY

### A. Data Collection

This study used raw data gathered from Twitter during an 8 month period between 1 July 2021 and 3 March 2022 from 7 major youth sustainability organizations and their corresponding network. Using Twitter's public API, relevant tweets along with user profiles from these 7 organizations were compiled and analyzed throughout the study period.

---

[18] Torfi, Amirsina, Rouzbeh A. Shirvani, Yaser Keneshloo, Nader Tavaf, and Edward A. Fox. "Natural language processing advancements by deep learning: A survey." *arXiv preprint arXiv:2003.01200* (2020).

[19] Tenney, Ian, Dipanjan Das, and Ellie Pavlick. "BERT rediscovers the classical NLP pipeline." *arXiv preprint arXiv:1905.05950* (2019).

[20] Murthy, Dhiraj, Alexander Gross, and Alexander Pensavalle. "Urban social media demographics: An exploration of Twitter use in major American cities." *Journal of Computer-Mediated Communication* 21, no. 1 (2016): 33-49.

Twitter's real-time filter stream was utilized, which allows users to select a list of people, keywords, and places to track. Four filter streams were developed to collect Tweets which mentioned specific users, were posted by a specified user, contained a selected keyword, and were posted from a specified location (See Table 1).

Table 1 Data collection streams, timeframe, and number of collected tweets

| Stream | Description | Timeframe | Tweets |
|---|---|---|---|
| 1 | Keywords: YOUNGO, UK Youth Climate Coalition, SDSN Youth, SustainUS, Youth Climate Movement, UK Student Climate Network, Southern Africa Youth Forum, GLF Youth, etc. | 1 July- 3 March | 30,149 |
| 2 | Accounts: YOUNGO, UKYCC, SDSN, SustainUS, COY, UKSCN1, GLF, etc. | 1 July- 3 March | 13,493 |
| 3 | 112 keywords related to youth sustainable development | 1 April- 1 March | 187,329 |
| 4 | Tweets geolocated in 7 youth conference locations | 48-hour period prior to event | 139,602 |

**B. Data Cleaning**

To decrease the noise in the social media data, all accounts, posts, and tweets which had no engagement in terms of likes, retweets, and comments were filtered out of the dataset. Additionally, retweets and comments which would connect a user to another account were used to connect the two users or nodes using edges in a graph. Prior research has displayed the connection between retweet networks, political ideology, and election results. [21] [22] Using similar techniques, a retweet and comment network where Twitter users were categorized into groups centred around the most retweeted or replied to twitter user. To filter out Twitter bots and irrelevant topic discussions, topics not containing words related to the keywords or the influential twitter user of the group were filtered out for manual review before being removed. Using a label propagation algorithm for community detection in social networks based on influence [23] [24], all groups with over 120 individuals were stored. The 3 communities with over 120 active youth users were identified: the largest was centered around United Nation Agencies (e.g. World Health

---

[21] Conover, Michael D., Bruno Gonçalves, Jacob Ratkiewicz, Alessandro Flammini, and Filippo Menczer. "Predicting the political alignment of twitter users." In *2011 IEEE third international conference on privacy, security, risk and trust and 2011 IEEE third international conference on social computing*, pp. 192-199. IEEE, 2011.

[22] Tumasjan, Andranik, Timm Sprenger, Philipp Sandner, and Isabell Welpe. "Predicting elections with twitter: What 140 characters reveal about political sentiment." In *Proceedings of the International AAAI Conference on Web and Social Media*, vol. 4, no. 1, pp. 178-185. 2010.

[23] Wu, Zhi-Hao, You-Fang Lin, Steve Gregory, Huai-Yu Wan, and Sheng-Feng Tian. "Balanced multi-label propagation for overlapping community detection in social networks." *Journal of Computer Science and Technology* 27, no. 3 (2012): 468-479.

[24] Zhang, Xian-Kun, et al. "Label propagation algorithm for community detection based on node importance and label influence." *Physics Letters A* 381.33 (2017): 2691-2698.

Organization), second was the Global Landscapes Forum (GLF), and third largest was the Children and Youth constituency to United Nations Framework Convention on Climate Change (YOUNGO). Combined, these communities compromised of over 18,305 youth users and over 308,403 tweets.

**C. Identifying Influential Users, User Demographics, and Topics from Twitter Data**

By creating a retweet and comment network where Twitter users were categorized into groups centred around the most retweeted or replied to twitter user, influential users could be identified by the PageRank algorithm. [25] [26] Each user in the dataset was given a PageRank score which determines the relative influence and authority from a scale of 0 to 10. The score is proportional to the chance of randomly reaching a node in the retweet and comment network. Thus, users with larger PageRank scores are more interconnected and have greater engagement than those with lower scores, identifying them as more influential and "authoritative" (See Table 2).

Using the pigeo Python library tweets were geolocated by country. [27] To obtain gender and age data, a computer vision approach was taken. A computer vision model pre-trained on the IMDB-WIKI dataset of 100,000 images of faces labeled by gender, age, and race. The validation-error for gender was 8% and the validation-error for age was 6.38 years. Tweets from users without a face in their profile picture and users with multiple faces in their profile picture did not have age or gender data. Users' first names and last names were contrasted with the U. S. Census data and were used as input for a text-based natural language processing recurrent neural network classier which infers gender and race based on historical data. Race was categorized into 5 categories: Asian, Hispanic, African, White, and Inconclusive. The Asian category encompassed people who are Indian, Japanese, and East Asian. The Hispanic category contained people whose names appeared in the Latino registered voters list, as well as the names which were classified as Hispanic European. African category contained people who in se names appeared in the Black registered voters list, as well as the names which were classified as Greater African. The White category encompassed people who had European names (except Hispanic Europeans) and people whose names appeared on the white registered voters list. This method of incorporating U. S. Census Bureau information with the text classifier has been an effective method for researchers in the health care setting. [28] Computer vision based approaches to infer race had less accuracy than name-based approaches in testing, so the text-based approach is employed.

---

[25] Wang, Rui, Weilai Zhang, Han Deng, Nanli Wang, Qing Miao, and Xinchao Zhao. "Discover community leader in social network with PageRank." In *International conference in Swarm intelligence*, pp. 154-162. Springer, Berlin, Heidelberg, 2013.

[26] Heidemann, Julia, Mathias Klier, and Florian Probst. "Identifying key users in online social networks: A pagerank based approach." (2010).

[27] Rahimi, Afshin, Trevor Cohn, and Timothy Baldwin. "pigeo: A python geotagging tool." In *Proceedings of ACL-2016 System Demonstrations*, pp. 127-132. 2016.

[28] Wong, Eric C., Latha P. Palaniappan, and Diane S. Lauderdale. "Using name lists to infer Asian racial/ethnic subgroups in the healthcare setting." *Medical care* 48, no. 6 (2010): 540.

Twitter user accounts without proper nouns in their name were filtered out. Additionally, users with no profile picture, ambiguous profile pictures or profile pictures with multiple individuals were removed. Tweets from users estimated to be aged over 25 were also filtered out. This resulted in a dataset with 11,406 twitter accounts and over 102,948 tweets.

Tweet content data was cleaned using "ekphrasis", a python library which utilizes a recurrent neural network trained on over one billion tweets to correct spelling and grammatical errors. Using the Universal Sentence Encoder (USE), a pre-trained transformer based model, along with the K-Nearest-Neighbor algorithm, tweet data was clustered and tested by reading random stratified subsamples of 30 topics. With *k* set to 250, the model performed best based upon manual validation techniques. All analysis was performed in python 3.6 and TensorFlow 2.9.

Table 2 Most Retweeted and Influential Accounts for Youth Twitter Users

| Rank | Retweeted | PageRank | Tweet Volume |
|---|---|---|---|
| 1 | UN Climate Change | UN | BBCWorld |
| 2 | UNFCCC | UN Climate Change | WorldBank |
| 3 | UN | WorldBank | UN |
| 4 | BBCWorld | Individual | UN Human Rights |
| 5 | Erik Solheim | UNYouth | estherclimate |

### III. RESULTS

#### A. Influential Accounts
In order to maintain privacy, individual accounts which are not representing public figures, governmental organizations, nonprofits, or corporations have been labelled as "Individual" (Table 2). Most of the 5 highest ranked twitter accounts for influence of youth twitter users are international organizations (e.g. UNFCCC, UN, UN Climate Change, UN Human Rights, BBCWorld, UN Human Rights, etc.). UNYouth being categorized amongst the five most influential accounts for youth twitter users based on PageRank displays the PageRank algorithm's effectiveness, as UNYouth has substantially less followers than twitter accounts run by organizations (ie. BBCWorld, UNFCCC, etc.). Additionally, individuals ranked amongst the top ten most influential accounts for youth twitter users using PageRank, included multiple youth leaders and a youth activist displaying accounts run by individuals also have influence over youth twitter users.

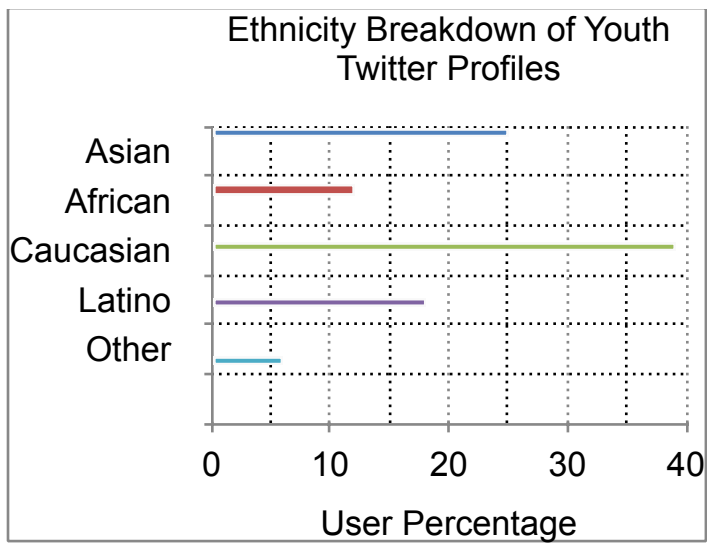

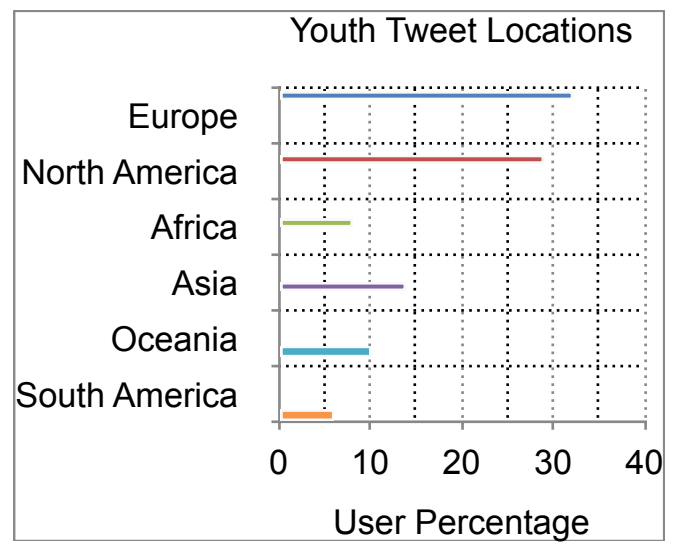

**Fig. 1** Continental distributions of studied youth Twitter users during 8 month study period

**Fig. 2** Ethnicity distributions of studied youth Twitter users during 8 month study period determined by RNN classifier.

## B. Topic Engagement Based on Demographics

Youth participation and involvement in topics which tended to go viral such as "#TeamSeas" and Political Unrest was disproportionately higher than other topics. Youth representation in topics such as World Bank Statistics and BBC World Breaking News was disproportionately less than other topics (Fig. 2). Young women were disproportionately more likely than young men to engage with tweets which were about topics such as: Animal rights, Food security, and Climate Change. They were disproportionately less likely than young men to engage with tweets which were about topics such as: Statistics and Data, Agriculture, and Corruption in Government. Caucasian people were disproportionately more likely than other races to engage with topics such as the United Nations, Data and Statistics, and Policy. They were disproportionately less likely than other races to engage with Industrial Pollution, Women's Rights, and Food Security. Additionally, Caucasians constituted a large majority of youth twitter profiles engaging in and participating in youth events and programs. Africans were a significant minority of youth twitter profiles engaging in youth events and programs for sustainable development.

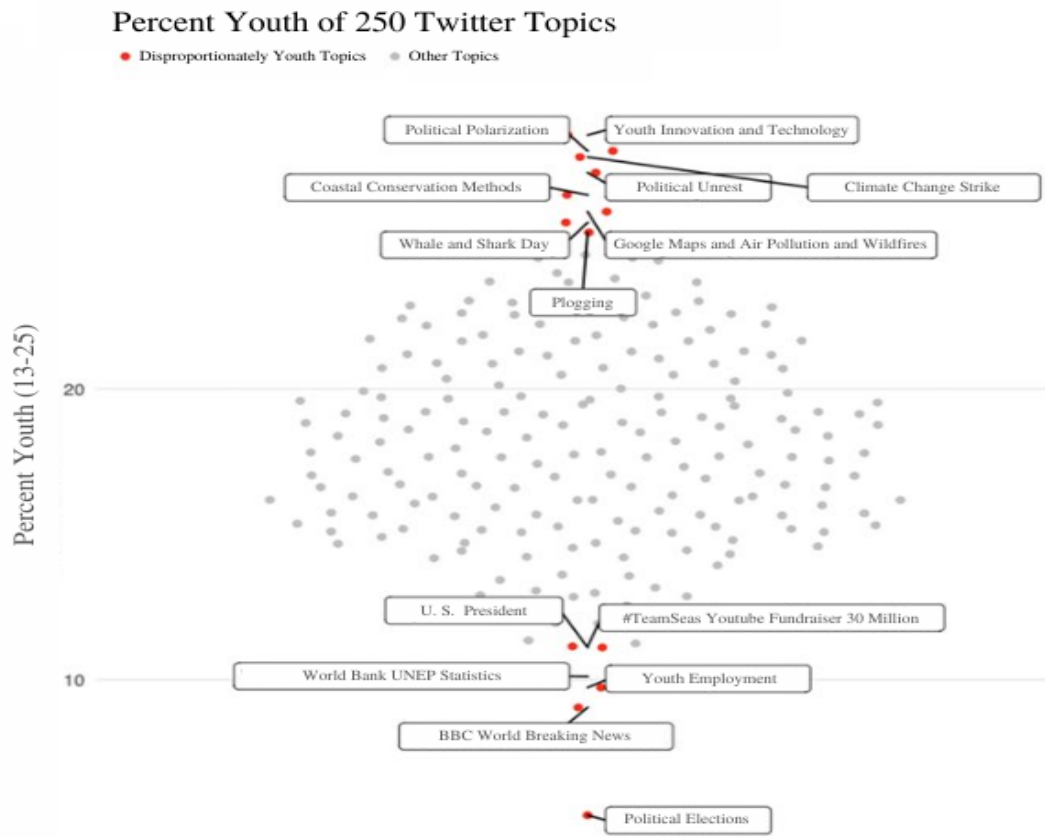

**Fig. 3** Topics related to study with disproportionate youth representation.

## IV. CONCLUSION

This work demonstrates how social media analysis using natural language processing can aid researchers in realizing how inclusive certain sustainable development programs and events are. Additionally, this study shows social media analysis provides an effective way to find sustainable development topics which disproportionately engage specific demographics, and this work shows how social media analysis can effectively find well-connected users who have large amounts of influence over certain demographics. This work highlighted the disproportionate engagement of certain demographics (e.g. Caucasian) as well as the topics which disproportionately engage and disengage the youth who are critical to the success of the Sustainable Development Goals. If organizations aiming to leverage youth support and input utilize these methods and appeal to lesser represented youth demographics, youth participation in achieving the 17 SDGs will increase in unprecedented ways.